\documentclass[10pt]{iopart}

\usepackage{iopams}  
\expandafter\let\csname equation*\endcsname\relax
\expandafter\let\csname endequation*\endcsname\relax
\usepackage{amsmath,amssymb}

\usepackage{tabularx} % for wide table
\usepackage{comment}

\usepackage{graphicx}% Include figure files
\usepackage{bm}% bold math
\usepackage{physics}
\usepackage[dvipsnames]{xcolor}
\usepackage[
    colorlinks,
    citecolor=RoyalBlue,
    linkcolor=RoyalBlue,
    urlcolor=RoyalBlue,
]{hyperref}
\usepackage{mathtools}
\usepackage{ dsfont }

\usepackage{fontawesome}

\newcommand{\red}[1]{\textcolor{Red}{#1}}

\newcommand{\green}[1]{\textcolor{Green}{#1}}

\usepackage{physics}
\usepackage{multirow}

\newcommand{\YES}{\green{\faCheck}\:}
\newcommand{\NO}{\red{\faClose}\:}

\newcommand{\boldsf}[1]{\textsf{\textbf{#1}}}

\begin{document}

\title{Scattering matrix for chiral harmonic generation and frequency mixing in nonlinear metasurfaces}

\author{Kirill~Koshelev$^{1,\dagger}$, Ivan~Toftul$^1$, Yongsop~Hwang$^2$, Yuri~Kivshar$^1$}
\ead{$^\dagger$kirill.koshelev@anu.edu.au}
\address{$^1$ Nonlinear Physics Center, Research School of Physics, Australian National University, Canberra, ACT 2601, Australia\\
$^2$ Laser Physics and Photonics Devices Lab, STEM, University of South Australia, Mawson Lakes, SA, 5095 Australia}

\vspace{10pt}
\begin{indented}
\item[]October 2023
\end{indented}

\begin{abstract}
We generalize the concept of optical scattering matrix ($S$-matrix) to characterize harmonic generation and frequency mixing in planar metasurfaces in the limit of undepleted pump approximation. We show that the symmetry properties of such nonlinear $S$-matrix are determined by the microscopic and macroscopic symmetries of the metasurface. We demonstrate that for description of degenerate frequency mixing processes such as optical harmonic generation, the multidimensional $S$-matrix can be replaced with a reduced two-dimensional $S$-matrix. We show that for metasurfaces possessing specific point group symmetries, the selection rules determining the transformation of the reduced nonlinear $S$-matrix are simplified substantially and can be expressed in a compact form. We apply the developed approach to analyse {\it chiral harmonic generation} in nonlinear metasurfaces with various symmetries including rotational, in-plane mirror, and out-of-plane mirror symmetries. For each of those symmetries, we confirm the results of the developed analysis by full-wave numerical calculations. We believe our results provide a new paradigm for engineering nonlinear optical properties of metasurfaces which may find applications in active and nonlinear optics, biosensing, and quantum information processing. 
\end{abstract}

\vspace{2pc}
\vspace{2pc}

\noindent This is the version of the article before peer review or editing, as submitted by the authors to {\it Journal of Optics}. IOP Publishing Ltd is not responsible for any errors or omissions in this version of the manuscript or any version derived from it.

\maketitle
\ioptwocol

\section{Introduction}
\label{sec:1}

Optical properties of metasurfaces composed of subwavelength planar metallic or dielectric elements have been attracting increased attention in the recent years~\cite{yu2014flat,chen2016review,hsiao2017fundamentals,koshelev2020dielectric,qiu2021quo}. Scattering properties of metasurfaces are substantially determined by the shape and arrangement of their meta-atoms. The description of the scattering processes can be conducted with the scattering matrix ($S$-matrix) formalism, which provides a relation between waves incoming at and outgoing from the metasurface. Breaking the symmetries of meta-atoms can be used for tailoring sharp resonant features in the scattering spectrum~\cite{koshelev2018asymmetric}, polarization control~\cite{pfeiffer2014high}, and other modifications of the transmitted and reflected wave amplitude and phase. The magnitude of these effects can be predicted by analysing the relative values of the elements of $S$-matrix. Numerically, $S$-matrix can be computed via different methods, including Fourier modal method~\cite{tikhodeev2002quasiguided}, decomposition into quasi-normal modes~\cite{gippius2005optical,weiss2018calculate, benzaouia2021quasi}, and temporal coupled-mode theory~\cite{fan2003temporal,zhang2020quasinormal}.

Enhancement of optical chirality and chiroptical effects in metasurfaces has been actively discussed in the past decade~\cite{oh2015chiral}. In contrast to natural chirality associated with a microscopic material structure of chiral ions and molecules, induced chiroptical response in metasurfaces arises because of broken geometrical symmetries of meta-atoms and their relative position. Chiroptical phenomena, indicative of optical activity, can be analysed through two distinct optical properties, optical rotation and circular dichroism. The latter describes of the unequal transmission or absorption of circularly polarized light within a chiral medium. Recently, the concept of optical $S$-matrix was utilized to derive the conditions of strong optical chirality and high circular dichroism depending on the symmetries of meta-atoms and their resonant properties~\cite{gorkunov2014extreme,kondratov2016extreme}. Specific properties of transmitted and reflected chiral light were linked to the metasurface symmetry, such as absence of polarization conversion in lattices with rotational symmetry~\cite{gorkunov2020metasurfaces}. Very recently, $S$-matrix was used to design single handedness cavities~\cite{voronin2022single}, and to achieve maximal optical chirality in metasurfaces via the engineering of coupling to sharp optical resonances~\cite{gorkunov2020metasurfaces,gorkunov2021bound}. 

Lately, the primary emphasis in optical metasurface research has transitioned towards nonlinear metasurfaces~\cite{minovich2015functional}.  In the last decade, it was shown that the nonlinear signal generated due to microscopic nonlinearities of metals or dielectrics can be enhanced strongly by engineering optical resonances~\cite{li2017nonlinear,keren2018shaping,krasnok2018nonlinear,koshelev2019nonlinear, liu2019high}. In particular, special interest was attracted to sum-frequency and optical harmonic generation in the undepleted pump regime~\cite{liu2016resonantly,camacho2022sum}, which goes beyond the phase-matching approach typical for nonlinear optics of macroscopic structures~\cite{armstrong1962interactions}. Very recently, it was shown that generation of higher-order harmonics can be achieved for low input power by virtue of sharp optical modes~\cite{shcherbakov2021generation,zograf2022high,tonkaev2023observation}.

Chiroptical activity in nonlinear metastructures and metasurfaces has been a topic of very intense studies~\cite{frizyuk2019second1,frizyuk2019second2,frizyuk2021nonlinear,nikitina2023nonlinear,koshelev2023nonlinear}. Analogous to linear phenomena, the enhancement of nonlinear chiroptical effects beyond the constraints of inherent material responses can be realized through breaking macroscopic symmetries of metasurface lattice and meta-atoms. It was shown that the efficiency of generation of chiral harmonics can be increased by employing resonances in metasurfaces~\cite{valev2009plasmonic,huttunen2011nonlinear,konishi2020circularly}, and maximal nonlinear chirality can be achieved~\cite{shi2022planar,koshelev2023resonant}. 

Pioneering studies of nonlinear chirality in optically active natural crystals and bulk media revealed that the nonlinear optical activity depends on the microscopic symmetry of susceptibility and permittivity tensors, and obeys a set of selection rules~\cite{simon1968second,alon1998selection}. Several later experimental studies showed that similar selection rules are also valid for nonlinear metasurfaces, meaning that only specific chiral harmonic generation processes are allowed with the restrictions imposed by the structure lattice symmetry~\cite{konishi2014polarization,chen2014symmetry,konishi2020circularly}. The question arises if a universal and simple approach to analyse nonlinear chiroptical response of metasurfaces can be constructed similar to the $S$-matrix calculations for linear metasurfaces.

In this paper, we generalize the concept of $S$-matrix to the description of nonlinear processes, such as frequency mixing, in the undepleted pump approximation [Eq.~\eqref{eq:Sdef}], schematically shown in Fig.~\ref{fig:1}. We derive the expression for the nonlinear $S$-matrix from Maxwell's equations [Eq.~\eqref{eq:appSfin}] that shows how the $S$-matrix is connected to the incoming and outgoing fields and material susceptibilities. We demonstrate that $S$-matrix symmetry properties are connected to the microscopic and macroscopic symmetries of the metasurface [Eq.~\eqref{eq:Str}]. We further show that for degenerate sum-frequency generation processes such as optical harmonic generation, the exact multidimensional $S$-matrix can be replaced with a reduced two-dimensional matrix [Eq.~\eqref{eq:SredN}]. We show that for metasurfaces possessing specific point symmetries [Eq.~\eqref{eq:Tcond}], the selection rules determining the transformation of the reduced nonlinear $S$-matrix are simplified substantially and can be expressed via multiplication of a small number of two-dimensional matrices [Eq.~\eqref{eq:SfinN}]. 

We apply the developed approach to analysis of selection rules for chiral harmonic generation in nonlinear metasurfaces excited in the normal incidence geometry. We analyse the fundamental and harmonic signals below the diffraction limit. We show that for metasurfaces with a rotational symmetry the harmonic signal generated by circularly polarized pump is allowed only in case of a specific relation between the harmonic order and symmetry order [Table~\ref{tb:2}]. We demonstrate that this relation originates from the conservation of the component of the total angular momentum of light projected on the out-of-plane axis of the metasurface. We show that depending on the harmonic order the allowed harmonics in reflection and transmission can be either only co- or only cross-polarized relative to the incident beam. We further study how in-plane mirror symmetries of the metasurface affect the chiral harmonic generation. We show that presence of an in-plane mirror symmetry leads to zero nonlinear circular dichroism for all harmonic orders [Table~\ref{tb:3}]. We next study the nonlinear chiroptical effects in metasurfaces with and without out-of-plane mirror symmetry. We show that for suspended metasurfaces with up-down reflection symmetry, the nonlinear circular dichroism is nonzero despite the structure becomes geometrically achiral [Table~\ref{tb:3}]. Finally, we derive  the selection rules for the case of three-dimensional inversion symmetry [Table~\ref{tb:3}]. For rotational and mirror refelction symmetries, we confirm the developed theory with full-wave numerical calculations. We believe the developed formalism paves a way towards a new generation of nonlinear metasurfaces which may find applications in active and nonlinear optics, biosensing and quantum information processing.

\section{Concept and symmetry selection rules}
\label{sec:2}

\subsection{$S$-matrix for linear scattering}

\begin{figure}
    \centering
    \includegraphics[width=0.85\linewidth]{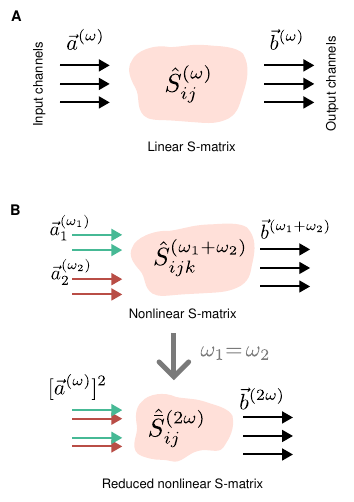}
    \caption{{\bf Concept of nonlinear optical $S$-matrix and its reduction for degenerate nonlinear processes}. \boldsf{A} Schematic of the conventional $S$-matrix for linear optical scattering in metasurfaces. Columns $\vec{a}^{(\omega)}$ and $\vec{b}^{(\omega)}$ describe the field amplitudes in the input and output scattering channels, respectively. In case the metasurface has a specific point symmetry, the $S$-matrix satisfies Eq.~\eqref{eq:SLtr}. 
    \boldsf{B} Top panel: Schematic of the nonlinear $S$-matrix for the specific case of sum-frequency generation with two input frequencies. We note that we assume the undepleted pump approximation. The nonlinear $S$-matrix is a matrix of the third rank in this case. We show that in case the metasurface has a specific point symmetry, the nonlinear $S$-matrix satisfies Eq.~\eqref{eq:Str}. Bottom panel: we introduce a concept of reduced nonlinear $S$-matrix described by a matrix of the second rank for degenerate sum-frequency generation, e.g. for ($\omega_1=\omega_2$), independently on the order of harmonic. For specific symmetry transformations, its elements obey selection rules in Eq.~\eqref{eq:SfinN0}. }
    \label{fig:1}
\end{figure}

We briefly recall the basic properties and definitions used for the conventional $S$-matrix describing linear scattering in planar metasurfaces. The $S$-matrix provides a relation between incoming and outgoing channels of the metasurface, as shown in Fig.~\ref{fig:1}A. A channel is identified as a solution to Maxwell's equations within the medium external to the metasurface. We treat this medium as two half-spaces composed of uniform and isotropic materials, separated by a flat interface. For such cases, Maxwell’s equations provide complete sets of orthogonal plane waves functions that can be used to expand an arbitrary solution of Maxwell’s equations outside of metasurface and on its surface. We provide explicit channel functions definition in \ref{app:1}.

For the transversal waves propagating normal to the metasurface plane, the $S$-matrix $\hat{S}^{(\omega)}$, schematically shown in Fig.~\ref{fig:1}A,  can be defined as
\begin{equation}
b_i^{(\omega)}=\sum\limits_{j}S^{(\omega)}_{ij}a_j^{(\omega)}.
\label{eq:SLdef1}
\end{equation}
Here, $i,j$ are the indices of the radiation channels, and $a_i$ and $b_i$ represent a set of incoming and outgoing amplitudes, respectively. Below the diffraction limit, the number of channels is limited to $4$, including two orthogonal porarizations and two directions of incidence (from upper side and from bottom side). Thus, the $S$-matrix can be written as a ($4\times4$) matrix of the second rank
\begin{equation}
\hat{S}^{(\omega)}=
\left(\begin{array}{cccc}
S_{11} & S_{12} & S_{13} & S_{14} \\
S_{12} & S_{22} & S_{23} & S_{24} \\
S_{13} & S_{23} & S_{33} & S_{34} \\
S_{14} & S_{24} & S_{34} & S_{44}
\end{array}\right).
\label{eq:SLdef2}
\end{equation}
%%%%%%%
\begin{figure*}[t]
    \centering
    \includegraphics[width=0.7\linewidth]{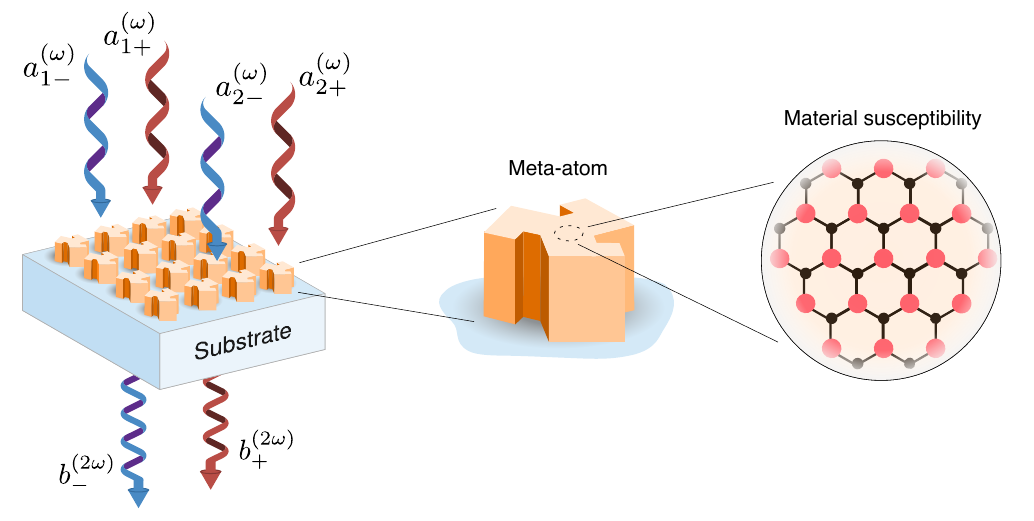}
    \caption{{\bf Nonlinear $S$-matrix for sum-frequency generation in metasurfaces}. We consider that the metasurface is excited in the normal incidence geometry. For each input and output frequency, we consider four channels numbered by two orthogonal polarizations and two directions of incidence (only two channels per frequency are displayed), provided we analyse only the zero-order diffraction normal to the metasurface plane. The symmetry of nonlinear $S$-matrix is determined by the point symmetry achieved at the macroscopic (metasurface lattice point symmetry, meta-atom point symmetry) and microscopic (permittivity and nonlinear susceptibility tensor point symmetry) scales.}
    \label{fig:2}
\end{figure*}
%%%%%%%%%%%%
We assume that the metasurface possesses a specific point symmetry simultaneously at the macroscopic scale for the unit cell and meta-atom, and at the microscopic scale for the permittivity. We assume that the point symmetry is described by the operator $\hat{T}$. In the selected coordinate basis, the transformation is characterized by the matrix $T_{ij}$, and
\begin{equation}
\begin{aligned}
    &b_i^{(\omega)} \rightarrow \sum\limits_{j}T_{ij}b_j^{(\omega)},\\
    &a_i^{(\omega)} \rightarrow \sum\limits_{j}T_{ij}a_j^{(\omega)}.
    \label{eq:ab}
\end{aligned}    
\end{equation}
The crystallographic symmetry principle (also called Neumann's principle)~\cite{hermann1934tensoren} imposes that the $S$-matrix must be invariant with respect to the symmetry operations of the point group of the metasurface unit cell. We can write that the corresponding operators commute $\hat{S}^{(\omega)}\hat{T}=\hat{T}\hat{S}^{(\omega)}$. Thus~,
\begin{equation}
    S^{(\omega)}_{ij} = \sum\limits_{l,p=1}^4T^{-1}_{il}S^{(\omega)}_{lp}T_{pj},\quad \text{for }i,j=1\ldots4.
    \label{eq:SLtr}
\end{equation}
Applying Eq.~(\ref{eq:SLtr}) for specific symmetry transformations $\hat{T}$, one can find relations between different elements of $\hat{S}^{(\omega)}$. This approach is widely used to analyse the conditions required for observation of strong chiroptical effects, e.g. in Refs.~\cite{gorkunov2014extreme,gorkunov2020metasurfaces}.

\subsection{$S$-matrix for nonlinear sum-frequency generation}

In this section, we aim to show that for specific symmetry transformations $\hat{T}$ and selected nonlinear processes of weak intensity in metasurfaces, we can introduce a reduced $(4\times4)$ nonlinear $S$-matrix, as schematically shown in Fig.~\ref{fig:1}B. The reduced matrix $\hat{\bar{S}}^{(N\omega)}$ fully describes the nonlinear process and obeys simple selection rules resembling Eq.~\eqref{eq:SLtr}, 
\begin{equation}
    \bar{S}^{(N\omega)}_{ij} = \sum\limits_{l,p=1}^4T^{-1}_{il}\bar{S}^{(N\omega)}_{lp}\left(T_{pj}\right)^N, \ \text{for }i,j=1
    \ldots4.
    \label{eq:SfinN0}
\end{equation}

We further derive Eq.~\eqref{eq:SfinN0} and show the limits of its applicability. We start with the general concept of $S$-matrix for nonlinear frequency mixing processes of weak intensity. We assume that the metasurface is pumped with $N$ multiple frequencies $\omega_1,\ldots,\omega_N$ and generates a signal at the output frequency $\omega_1+\ldots+\omega_N$, as shown in the left panel of Fig.~\ref{fig:2}. We also assume the perturbative regime and undepleted pump approximation for the frequency conversion, which is widely used for dielectric and plasmonic metasurfaces due to low conversion efficiency. Then, the nonlinear $S$-matrix, schematically shown in the top panel of Fig.~\ref{fig:1}B, can be defined as
\begin{equation}
b_i^{(\omega_1+\ldots+\omega_N)}=\sum\limits_{j,\ldots,k}S^{(\omega_1+\ldots+\omega_N)}_{ij\dots k}a_j^{(\omega_1)}\cdot\ldots\cdot a_k^{(\omega_N)}.
\label{eq:Sdef}
\end{equation}
Here, the nonlinear $S$-matrix is a matrix of the $n$-th rank, and $i,j,\ldots,k$ are the channel indices as in Eq.~\eqref{eq:SLdef1}. 

We can analyse the symmetry properties of $\hat{S}^{(\omega_1+\ldots+\omega_N)}$ for metasurfaces characterized with a certain point group symmetry element $\hat{T}$ by combining Eqs.~\eqref{eq:ab} and \eqref{eq:Sdef}, resulting in
\begin{equation}
    S^{(\omega_1+\ldots+\omega_N)}_{ij\ldots k} = \sum\limits_{l,\ldots,p,s}T^{-1}_{il}S^{(\omega_1+\ldots+\omega_N)}_{lp\ldots s}T_{pj}\cdot\ldots\cdot T_{sk}.
    \label{eq:Str}
\end{equation}
Importantly, $\hat{T}$ is an element of the point group symmetry of the nonlinear $S$-matrix, only if it is included in the point symmetry group elements of the metasurface lattice, meta-atom, and material susceptibility tensors (Fig.~\ref{fig:2}). The proof is given in \ref{app:1}. We also notice that the proposed nonlinear $S$-matrix is non-reciprocal, and does not obey energy conservation and time-reversal symmetry by definition, because we consider the undepleted pump approximation.

One can see that the analysis of Eq.~(\ref{eq:Str}) is challenging due to a large dimensionality of the nonlinear $S$-matrix. Our goal is to show that for selected simple sum-frequency generation processes, such as generation of optical harmonics with a polarized pump beam, only specific elements of $S^{(\omega_1+\ldots+\omega_N)}_{ij\dots k}$ are needed to describe the process, and Eq.~(\ref{eq:Sdef}) can be simplified to a form similar to Eq.~(\ref{eq:SLtr}). 

\paragraph{Second-harmonic generation.} We next consider the second-harmonic generation (SHG) process, for the sake of clarifity of the derivation. We note that the model developed below can be generalized for higher-order harmonic generation. For SHG with the pump and second-harmonic (SH) wavelength below the diffraction limit, the number of open radiative channels is $4$, harmonic order $N=2$, and $\omega_1=\omega_2=\omega$. Therefore,
\begin{equation}
    {a}^{(\omega_1)}_i={a}^{(\omega_2)}_i,\quad \text{for }i=1\ldots4.
    \label{eq:a1a2}
\end{equation}
The nonlinear $S$-matrix $\hat{S}^{(2\omega)}$ for SHG is a matrix of the third rank with the shape of ($4\times4\times4$). Its definition can be written as [see Eq.~(\ref{eq:Sdef})] 
\begin{equation}
b_i=\sum\limits_{j,k=1}^4S^{(2\omega)}_{ijk}a_ja_k.
\label{eq:newSdef}
\end{equation}
We note that from here on we omit the frequency superscript for amplitudes $a_i$ and $b_i$. 

In Eq.~(\ref{eq:newSdef}), the terms with $S^{(2\omega)}_{ijk}$ and $S^{(2\omega)}_{ikj}$ can be combined because of Eq.~(\ref{eq:a1a2}). Then, we can re-write Eq.~(\ref{eq:newSdef}) in a simpler form
\begin{equation}
\left(\begin{array}{c}
b_1 \\
b_2 \\
b_3 \\
b_4
\end{array}\right)=\hat{\mathbb{S}}^{(2\omega)}\left(\begin{array}{c}
\left(a_1\right)^2 \\
\left(a_2\right)^2 \\
\left(a_3\right)^2 \\
\left(a_4\right)^2 \\
2a_1a_2 \\
2a_1a_3 \\
2a_1a_4 \\
2a_2a_3 \\
2a_2a_4 \\
2a_3a_4 \\
\end{array}\right).
\label{eq:Sbig}
\end{equation}
Here, we introduce an auxiliary $S$-matrix  $\hat{\mathbb{S}}^{(2\omega)}$, which is a matrix of the second rank with the shape of ($4\times10$) and elements defined as
\begin{equation}
    {\mathbb{S}}^{(2\omega)}_{i\alpha}\equiv\frac{{S}^{(2\omega)}_{ijk}+{S}^{(2\omega)}_{ikj}}{2}, \quad \text{for }\alpha(j,k)=1\ldots10.
    \label{eq:Saux2}
\end{equation}
From here on, we use Greek symbols $\alpha,\beta,\ldots$ for the indices ranging from $1$ to $10$, and Roman symbols $i,j,k,\ldots$ for the indices in the range from $1$ to $4$. The correspondence between indexing with $\alpha$ and $(j,k)$ is shown in Table~\ref{tb:1}. For example,  ${\mathbb{S}}^{(2\omega)}_{12}={S}^{(2\omega)}_{122}$, ${\mathbb{S}}^{(2\omega)}_{35}=({S}^{(2\omega)}_{312}+{S}^{(2\omega)}_{321})/2$. We note the number of columns of ${\mathbb{S}}^{(2\omega)}_{i\alpha}$ is defined as $\alpha_{\rm max}=d(d+1)/2$, where $d$ is the number of open channels (see more details in~\ref{app:2}). In our case, $d=4$ and $\alpha_{\rm max}=10$.

\begin{table}[t]
    \centering
    \caption{\label{tb:1} Correspondence between indices $\alpha$ and $(j,k)$.}
    \begin{tabularx}{\linewidth}{XX|XX} % Use tabularx and set the width to linewidth
    %\hline \hline
    \mr
    $\alpha$& $(j,k)$ & $\alpha$& $(j,k)$\\
    %\hline 
    \mr
    $1$&$(1, 1)$& $6$&$(1, 3)$\\
    $2$&$(2, 2)$& $7$&$(1, 4)$\\
    $3$&$(3, 3)$& $8$&$(2, 3)$\\
    $4$&$(4, 4)$& $9$&$(2, 4)$\\
    $5$&$(1, 2)$& $10$&$(3, 4)$\\
    %\hline \hline
    \mr
    \end{tabularx}
\end{table}
\begin{comment}
\begin{table}[t]
\centering
\caption{\label{tb:1} Correspondence between indices $\alpha$ and $(j,k)$.}
\footnotesize
\begin{tabular}{|c|c||c|c|}
\mr
$\alpha$& $(j,k)$ & $\alpha$& $(j,k)$\\
\mr
$1$&$(1, 1)$& $6$&$(1, 3)$\\
$2$&$(2, 2)$& $7$&$(1, 4)$\\
$3$&$(3, 3)$& $8$&$(2, 3)$\\
$4$&$(4, 4)$& $9$&$(2, 4)$\\
$5$&$(1, 2)$& $10$&$(3, 4)$\\
\mr
\end{tabular}\\
\end{table}
\normalsize
\end{comment}

The $S$-matrix $\hat{S}^{(2\omega)}$ obeys the selection rules governed by [see Eq.~(\ref{eq:Str})] 
\begin{equation}
    S^{(2\omega)}_{ijk} = \sum\limits_{l,p,s}T^{-1}_{il}S^{(2\omega)}_{lps}T_{pj}T_{sk}.
\label{eq:StrSHG}
\end{equation}
We can re-write Eq.~(\ref{eq:StrSHG}) in terms of transformation of $\hat{\mathbb{S}}^{(2\omega)}$ by combining the corresponding elements of $\hat{T}$,
\begin{equation}
    {\mathbb{S}}^{(2\omega)}_{i\alpha} = \sum\limits_{l=1}^4\sum\limits_{\beta=1}^{10}T^{-1}_{il}{\mathbb{S}}^{(2\omega)}_{l\beta}\mathbb{T}^{(2)}_{\beta\alpha}.
    \label{eq:StrT}
\end{equation}
Here, $\hat{\mathbb{T}}^{(2)}$ is the auxiliary symmetry transformation operator with the matrix of ($10\times10$) shape defined as
\begin{equation}
    \mathbb{T}^{(2)}_{\alpha\beta}\equiv\frac{T_{jp}T_{ks}+T_{js}T_{kp}}{2}, \quad  \text{for }\alpha(j,k),\ \beta(p,s).
\end{equation}
The correspondence between $\beta=1\ldots10$ and $p,s=1\ldots4$ is the same as between $\alpha$ and $(j,k)$ (see Table~\ref{tb:1}). For example,  ${\mathbb{T}}^{(2)}_{12}=T_{12}^2$, ${\mathbb{T}}^{(2)}_{35}=T_{31}T_{32}$, ${\mathbb{T}}^{(2)}_{78}=(T_{12}T_{43}+T_{13}T_{42})/2$.

We focus on incident beams with well-defined polarization, such as linearly- or circularly polarized plane waves. With a proper choice of coordinate basis for $\mathbf{b}$ and $\mathbf{a}$ matching the incident wave polarization, we can impose $a_ia_j=0$ for $(i\ne j)$ and further simplify Eq.~(\ref{eq:StrT}). In this case, the input amplitude has a single nonzero component, so the output amplitudes $b_i$ are defined only by the elements ${\mathbb{S}}^{(2\omega)}_{i\alpha}$ with $\alpha=1\ldots4$, as can be seen from Eq.~(\ref{eq:Sbig}). We can then define this part of $\hat{\mathbb{S}}^{(2\omega)}$ as a reduced nonlinear $S$-matrix $\hat{\bar{S}}^{(2\omega)}$
\begin{equation}
\begin{aligned}
&\bar{S}^{(2\omega)}_{ij}\equiv{\mathbb{S}}^{(2\omega)}_{i,\alpha(j,j)}={S}^{(2\omega)}_{ijj}, \quad \text{for }i,j=1\ldots4,\\
&b_i=\sum\limits_{j=1}^4\bar{S}^{(2\omega)}_{ij}\left(a_j\right)^2.
\label{eq:Seffdef}
\end{aligned}
\end{equation}
The reduced $S$-matrix $\hat{\bar{S}}^{(2\omega)}$ is a matrix of the second rank with the shape of ($4\times4$) similar to the $S$-matrix $\hat{S}^{(\omega)}$ used for description of linear scattering.
The selection rules for the reduced nonlinear $S$-matrix are defined via
\begin{multline}
    \bar{S}^{(2\omega)}_{ij} = \sum\limits_{l,p=1}^4T^{-1}_{il}\bar{S}^{(2\omega)}_{lp}\mathbb{T}^{(2)}_{\beta(p,p),\alpha(j,j)}+\\
    +\sum\limits_{l=1}^4\sum\limits_{\beta=5}^{10}T^{-1}_{il}{\mathbb{S}}^{(2\omega)}_{l\beta}\mathbb{T}^{(2)}_{\beta,\alpha(j,j)}.
    \label{eq:Spprfin}
\end{multline}
The first term in Eq.~(\ref{eq:Spprfin}) is defined via the reduced $S$-matrix only, resembling Eq.~(\ref{eq:SLtr}), while the second term cannot be expressed via $\hat{\bar{S}}^{(2\omega)}$. The transformation matrix in the second term can be written as $\mathbb{T}^{(2)}_{\beta(p,s),\alpha(j,j)}=T_{pj}T_{sj}$ with $j,p,s=1\ldots4$ and $p\ne s$ due to $\beta(p,s)=5\ldots10$.

\begin{figure}[t]
    \centering
    \includegraphics[width=0.70\linewidth]{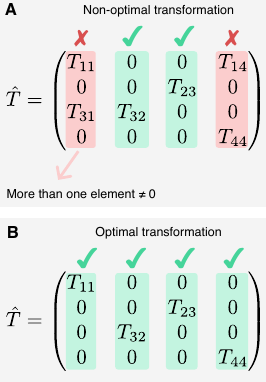}
    \caption{{\bf Optimal and non-optimal transformation matrices}. \boldsf{A} General form of a non-optimal transformation matrix $\hat{T}$. \boldsf{B} General form of an optimal transformation matrix $\hat{T}$. The optimal criteria are satisfied the transformation matrix does not mix multiple channels, i.e. if each column of $\hat{T}$ has only one non-zero element. The selection rules analysis is greatly simplified for optimal matrices, as Eq.~\eqref{eq:Spprfin} is transformed to Eq.~\eqref{eq:Sfin2} that contains two-dimensional matrices only. In the basis of circularly polarized waves with defined helicity, the optimal criteria are satisfied for all relevant point group symmetries, including the rotational, in-plane mirror, out-of-plane mirror and inversion symmetries.}
    \label{fig:3}
\end{figure}

We focus on specific symmetry transformations obeying $T_{pj}T_{sj}=0$ for all $p\ne s$ in the given coordinate basis. This means that the each column of the matrix $T_{pj}$ has no more than one nonzero element, as schematically shown in Fig.~\ref{fig:3}. Such transformations do not mix multiple channels, according to Eq.~\eqref{eq:ab}. As shown below in Sec.~\ref{sec:3}, this condition is valid for a large set of symmetries, especially, in the basis of circularly polarized waves with defined helicity. For this condition, the second term in Eq.~(\ref{eq:Spprfin}) is zero, and, using $\mathbb{T}^{(2)}_{\beta(p,p),\alpha(j,j)}=T_{pj}^2$, we can write
\begin{equation}
    \bar{S}^{(2\omega)}_{ij} = \sum\limits_{l,p=1}^4T^{-1}_{il}\bar{S}^{(2\omega)}_{lp}T_{pj}^2.
    \label{eq:Sfin2}
\end{equation}

\paragraph{Arbitrary order of harmonic.} For harmonic generation of an arbitrary order $N$, we can introduce a reduced $(4\times4)$ nonlinear $S$-matrix in analogy to Eq.~(\ref{eq:Seffdef}),
\begin{equation}
\begin{aligned}
    &\bar{S}^{(N\omega)}_{ij} \equiv{S}^{(N\omega)}_{ij\dots j},\quad \text{for } i,j=1\ldots4\\ &b_i^{(N\omega)}=\sum\limits_{j=1}^4\bar{S}^{(N\omega)}_{ij}\left(a_j^{(\omega)}\right)^N.
\end{aligned}
\label{eq:SredN}
\end{equation}
Following the steps above (see \ref{app:2} for details), one can show that $\hat{\bar{S}}^{(N\omega)}$ obeys 
\begin{equation}
    \bar{S}^{(N\omega)}_{ij} = \sum\limits_{l,p=1}^4T^{-1}_{il}\bar{S}^{(N\omega)}_{lp}\left(T_{pj}\right)^N.
    \label{eq:SfinN}
\end{equation}
under specific symmetry transformations $\hat{T}$ that follow
\begin{equation}
    T_{pj}T_{sj}=0, \quad \text{for } p,s=1\ldots4;\ p\ne s.
    \label{eq:Tcond}
\end{equation}
Equations~(\ref{eq:SfinN}) and (\ref{eq:Tcond}) are the central result of our study. They represent the generalization of selection rules given by Eq.~(\ref{eq:SLtr}) for harmonic generation processes in the undepleted pump approximation. We note that for ($N=1$) Eq.~\eqref{eq:SfinN} transforms to Eq.~(\ref{eq:SLtr}), as expected. Importantly, instead of analysing the multidimensional matrix ${S}^{(N\omega)}_{ij\ldots k}$, the selection rules can be predicted from multiplication of three two-dimensional ($4\times4$) matrices,  $T^{-1}_{ij}$, $\left(T_{ij}\right)^N$ and $\bar{S}_{ij}^{(N\omega)}$.

\section{Application to chiral harmonic generation and numerical validation}
\label{sec:3}
\subsection{Model and definitions}
We consider a metasurface shown schematically in the left panel of Fig.~\ref{fig:2}, located in the $(x-y)$ plane. We assume $\exp(-i\omega t)$ time dependence for all fields. We consider that the nonlinear $S$-matrix is expressed in the basic of plane waves with defined helicity, i.e. polarized along the complex unit vectors
\begin{equation}
\mathbf{e}_{ \pm}=\left(\mathbf{e}_x \mp i \mathbf{e}_y\right) / \sqrt{2}.
\end{equation}
For waves propagating along the positive $z$ direction, $\mathbf{e}_{+}$ and $\mathbf{e}_{-}$ correspond to the right circular polarization (RCP) and left circular polarization (LCP), respectively. For waves propagating along the negative $z$ direction, the sign of polarizations is flipped and $\mathbf{e}_{+}$ and $\mathbf{e}_{-}$ correspond to the LCP and RCP, respectively.

The generation of the $N$-th order optical harmonic can be described by the reduced $S$-matrix equation relating the outgoing wave amplitudes $b_{\pm,{\rm u(d)}}$ at the harmonic frequency $(N\omega)$ with the incident wave amplitudes $a_{\pm,{\rm u(d)}}$,
\begin{align}
&\left(\begin{array}{c}
b_{+,{\rm u}} \\
b_{+,{\rm d}} \\
b_{-,{\rm u}} \\
b_{-,{\rm d}}
\end{array}\right)=\hat{\bar{S}}^{(N\omega)}\left(\begin{array}{c}
\left(a_{+,{\rm u}}\right)^N \\
\left(a_{+,{\rm d}}\right)^N \\
\left(a_{-,{\rm u}}\right)^N \\
\left(a_{-,{\rm d}}\right)^N
\end{array}\right),\\
&\hat{\bar{S}}^{(N\omega)}=\left(\begin{array}{cccc}
r_{\rm LR, u}^{(N)} & t_{\rm LL, d}^{(N)} & r_{\rm LL, u}^{(N)} & t_{\rm LR, d}^{(N)} \\
t_{\rm RR, u}^{(N)} & r_{\rm RL, d}^{(N)} & t_{\rm RL, u}^{(N)} & r_{\rm RR, d}^{(N)} \\
r_{\rm RR, u}^{(N)} & t_{\rm RL, d}^{(N)} & r_{\rm RL, u}^{(N)} & t_{\rm RR, d}^{(N)} \\
t_{\rm LR, u}^{(N)} & r_{\rm LL, d}^{(N)} & t_{\rm LL, u}^{(N)} & r_{\rm LR, d}^{(N)}
\end{array}\right).
\label{eq:Shel}
\end{align}
Here, the parameters $t_{ij,{\rm u(d)}}^{(N)}$ and $r_{ij,{\rm u(d)}}^{(N)}$ determine forward and backward scattered $N$-th harmonic signal, respectively, with $i = {\rm L,R}$ polarization of the outgoing waves for $j = {\rm L, R}$ polarized pump, and $({\rm u,d})$ define "up" ($z<0$) and "down" ($z>0$) direction of incoming wave. From here on, we omit the subscript $\pm$ for the reduced $S$-matrix for the sake of clarity.
 %%%%%%%%%
\begin{figure}
    \centering
    \includegraphics[width=0.9\linewidth]{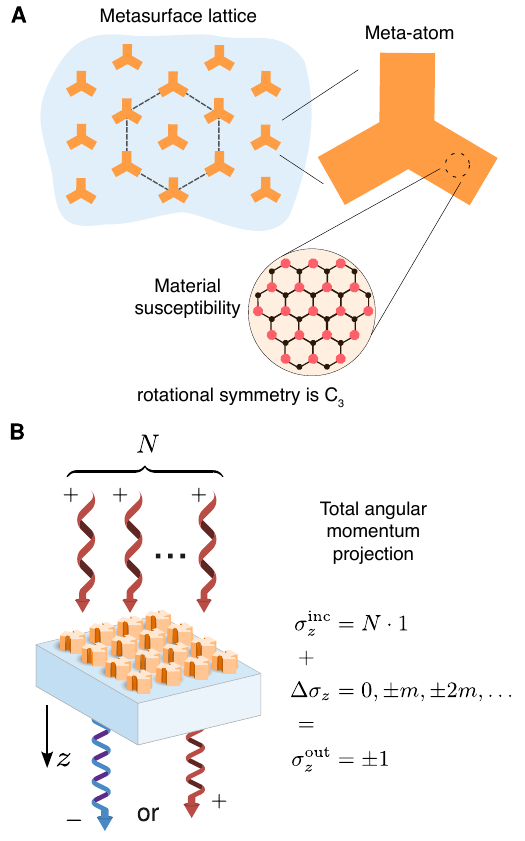}
    \caption{{\bf General concept for metasurfaces with rotational symmetries}. \boldsf{A} The point group symmetry of the nonlinear $S$-matrix consists of identical elements of the point symmetry groups of rotational symmetry of the metasurface lattice, meta-atom, and material susceptibility tensors. 
    \boldsf{B} Schematic of the total angular momentum projection conservation. The total angular momentum $\sigma_z^{\text{inc}}=N$ (in units of $\hbar$) of the incident beam composed of $N$ right-polarized degenerate waves is summed up with the $\Delta \sigma_z$ arising from the interaction with the metasurface. $\Delta \sigma_z$ is equal to an integer number of $m$ because the metasurface has an angular period related to the rotational symmetry of $m$-th order ${\rm C}_m$ and conserves total angular quasi-momentum. The total angular momentum per one emitted photon for the output beam at harmonic frequency must be equal to $\pm1$. If this condition could not be fulfilled, the chiral harmonic generation of specific order is forbidden.
    }
    \label{fig:4}
\end{figure}
%%%%%%%%%

We will focus on the key optical parameter describing chiroptical effects, circular dichroism (CD). We extend its definition to nonlinear harmonic generation. In general, due to nonlinear circular birefringence, an LCP wave at $(\omega)$ generates both LCP and RCP waves at $(N\omega)$. 

% We can define a co- and cross-polarized OR for $p={\rm u,d}$ as
% \begin{equation}
% \begin{aligned}
% \mathrm{OR}^{(N)}_{p,{\rm co}}&=\frac{1}{2}\left(\arg t_{{\rm LL}, p}^{(N)}-\arg t_{{\rm RR}, p}^{(N)}\right),\\
% \mathrm{OR}^{(N)}_{p,{\rm cross}}&=\frac{1}{2}\left(\arg t_{{\rm RL}, p}^{(N)}-\arg t_{{\rm LR}, p}^{(N)}\right).
% \label{eq:ORall}
% \end{aligned}
% \end{equation}
We can define co- and cross-polarized nonlinear CD for $p={\rm u,d}$,
\begin{equation}
\label{eq:CDall}
\begin{aligned}
\mathrm{CD}^{(N)}_{p,{\rm co}}&=\frac{\left|t_{{\rm RR}, p}^{(N)}\right|^2-\left|t_{{\rm LL}, p}^{(N)}\right|^2}{\left|t_{{\rm RR}, p}^{(N)}\right|^2+\left|t_{{\rm LL}, p}^{(N)}\right|^2},\\
\mathrm{CD}^{(N)}_{p,{\rm cross}}&=\frac{\left|t_{{\rm RL}, p}^{(N)}\right|^2-\left|t_{{\rm LR}, p}^{(N)}\right|^2}{\left|t_{{\rm RL}, p}^{(N)}\right|^2+\left|t_{{\rm LR}, p}^{(N)}\right|^2},
\end{aligned}
\end{equation}
and total nonlinear CD,
\begin{equation}
    \mathrm{CD}^{(N)}_{p, {\rm tot}}=\frac{\left|t_{{\rm RR}, p}^{(N)}\right|^2+\left|t_{{\rm LR}, p}^{(N)}\right|^2-\left|t_{{\rm LL}, p}^{(N)}\right|^2-\left|t_{{\rm RL}, p}^{(N)}\right|^2}{\left|t_{{\rm RR}, p}^{(N)}\right|^2+\left|t_{{\rm LL}, p}^{(N)}\right|^2+\left|t_{{\rm LL}, p}^{(N)}\right|^2+\left|t_{{\rm RL}, p}^{(N)}\right|^2}.
\end{equation}

To calculate the matrices $\hat{T}_{\pm}$ for specific symmetry transformations in the helical amplitude basis, we use a unitary transformation of corresponding matrices $\hat{T}_{x y}$ defined in the Cartesian amplitude basis. The matrices $\hat{T}_{x y}$ for various symmetry transformations are defined in \ref{app:3}.

\subsection{Rotational symmetry of $m$-th order}

We assume that the structure has a rotational symmetry of $m$-th order (${\rm C}_m$) at both microscopic and macroscopic scale, as schematically shown for the case of ${\rm C}_3$ in Fig.~\ref{fig:4}A. The matrix corresponding to ${\rm C}_m$ transformation is
\begin{equation}
\hat{T}^{(\varphi_m)}_{\pm}=\left(\begin{array}{cccc}
e^{i\varphi_m} & 0 & 0 & 0 \\
0 & e^{i\varphi_m}  & 0 & 0 \\
0 & 0 & e^{-i\varphi_m}  & 0 \\
0 & 0 & 0 & e^{-i\varphi_m} 
\end{array}\right),
\end{equation}
where $\varphi_m=2\pi/m$. The matrix is unitary, thus $\left(\hat{T}^{(\varphi_m)}_{\pm}\right)^{-1}=\hat{T}^{(-\varphi_m)}_{\pm}$. We can write the selection rules Eq.~\eqref{eq:SfinN} as
\begin{equation}
    \hat{\bar{S}}^{(N\omega)} = \hat{T}^{(-\varphi_m)}_{\pm}\hat{\bar{S}}^{(N\omega)}\hat{T}^{(N\varphi_m)}_{\pm}.
    \label{eq:Shelrot}
\end{equation}
Applying the selection rules to Eq.~(\ref{eq:Shelrot}), we get four distinctive cases. First, if we can find an integer $s_{1}$ that for any integer $s_2$
\begin{equation}
\begin{aligned}
N+ms_1=1,\quad N+ms_2\ne-1,
\end{aligned}
\end{equation}
we get the constraints for the coefficients,
\begin{equation}
\begin{aligned}
&r_{\rm LL, u(d)}^{(N)}=0,\quad r_{\rm RR, u(d)}^{(N)}=0,\\
&t_{\rm LR, u(d)}^{(N)}=0,\quad t_{\rm RL, u(d)}^{(N)}=0,
\end{aligned}
\label{eq:const1}
\end{equation}
and the nonlinear $S$-matrix takes a simplified form,
\begin{equation}
\hat{\bar{S}}^{(N\omega)}=\left(\begin{array}{cccc}
r_{\rm LR, u}^{(N)} & t_{\rm LL, d}^{(N)} & 0 & 0 \\
t_{\rm RR, u}^{(N)} & r_{\rm RL, d}^{(N)} & 0 & 0 \\
0 & 0 & r_{\rm RL, u}^{(N)} & t_{\rm RR, d}^{(N)} \\
0 & 0 & t_{\rm LL, u}^{(N)} & r_{\rm LR, d}^{(N)}
\end{array}\right).
\label{eq:Sform1}
\end{equation}
The CD calculation is simplified as
\begin{equation}
    \mathrm{CD}^{(N)}_{p, {\rm tot}} = \mathrm{CD}^{(N)}_{p, {\rm co}}.
\end{equation}
Second, if we can find an integer $s_{1}$ that for any integer $s_2$
\begin{equation}
\begin{aligned}
N+ms_1=-1,\quad N+ms_2\ne1,
\end{aligned}
\end{equation}
we get the constraints for the coefficients,
\begin{equation}
\begin{aligned}
&r_{\rm LR, u(d)}^{(N)}=0,\quad r_{\rm RL, u(d)}^{(N)}=0,\\
&t_{\rm LL, u(d)}^{(N)}=0,\quad t_{\rm RR, u(d)}^{(N)}=0,
\end{aligned}
\label{eq:const2}
\end{equation}
and the nonlinear $S$-matrix takes a simplified form,
\begin{equation}
\hat{\bar{S}}^{(N\omega)}=\left(\begin{array}{cccc}
0 & 0 & r_{\rm LL, u}^{(N)} & t_{\rm LR, d}^{(N)} \\
0 & 0 & t_{\rm RL, u}^{(N)} & r_{\rm RR, d}^{(N)} \\
r_{\rm RR, u}^{(N)} & t_{\rm RL, d}^{(N)} & 0 & 0 \\
t_{\rm LR, u}^{(N)} & r_{\rm LL, d}^{(N)} & 0 & 0
\end{array}\right).
\label{eq:Sform2}
\end{equation}
The CD calculation is simplified as
\begin{equation}
    \mathrm{CD}^{(N)}_{p, {\rm tot}} = \mathrm{CD}^{(N)}_{p, {\rm cross}}.
\end{equation}
Third, if we can find integers $s_{1}$ and $s_2$ that
\begin{equation}
\begin{aligned}
N+ms_1=1,\quad N+ms_2=-1,
\end{aligned}
\end{equation}
we do not have any additional constraints. Finally, if for any integer $s$
\begin{equation}
N+ms\ne\pm1,
\label{eq:22}
\end{equation}
all $S$-matrix elements are zero
\begin{equation}
    \hat{\bar{S}}^{(N\omega)}_{ij}= {0},\quad \text{for } i,j=1     \ldots4.
\end{equation}

Equations $N+ms=\pm1$ represent the conservation law for the projection of total angular momentum of light on the $z$ axis for chiral harmonic generation processes, shown schematically in Fig.~\ref{fig:4}B. We can interpret them as follows: $N$ photons with the total momentum projection $+1$ excite the metasurface. Since the metasurface lattice, meta-atom and molecular lattice all possess rotational symmetry of $m$-th order, the incident photons can acquire a multiple integer number $s$ of total momentum projection quanta $m$. The output photon at harmonic frequency can carry the total momentum projection $\pm1$, thus the emitted photons should obey the conservation law. If these conditions cannot be fulfilled, chiral harmonic generation is prohibited due to inconsistency with the total angular momentum projection conservation.
%%%%%%
\begin{figure*}[t]
    \centering
    \includegraphics[width=\linewidth]{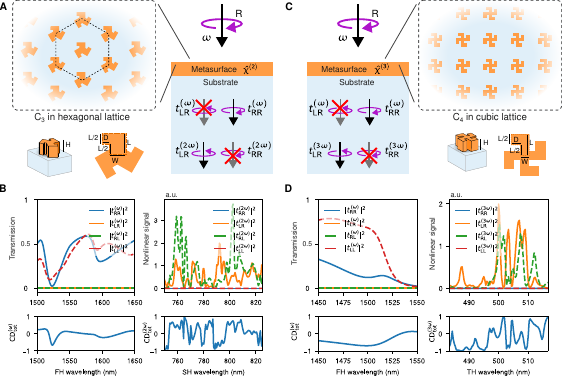}
    \caption{{\bf Numerical analysis for metasurfaces with rotational symmetries.}
    \boldsf{A},\boldsf{C} Schematic of linear and nonlinear transmission in metasurfaces with $\mathrm{C}_3$ meta-atoms of $\mathrm{C}_3$ material in a hexagonal lattice \boldsf{A} and for $\mathrm{C}_4$ meta-atoms of $\mathrm{C}_4$ material in a square lattice \boldsf{C}. \boldsf{B},\boldsf{D} Calculated linear and nonlinear transmission signals and CD. The pump wave is incident from the upper side, index $p={\rm u}$ is omitted.
    Parameters of the simulations for the metasurface in \boldsf{A}, \boldsf{B}: $L = 530~\text{nm}$, $D=180~\text{nm}$, $W=430~\text{nm}$, $H = 640~\text{nm}$, period $1440~\text{nm}$, substrate permittivity $3.13$, material permittivity $16.7$ at pump and $22.5$ at SH wavelength, respectively, $\hat{\chi}^{(2)}$ is defined by the point group 3m (e.g. MoS$_2$). Parameters of the simulations for the metasurface in \boldsf{C}, \boldsf{D}: $L = 550~\text{nm}$, $D=250~\text{nm}$, $W=230~\text{nm}$, $H = 400~\text{nm}$, period $1300$~\text{nm}, substrate permittivity $2$, material permittivity $16$ at pump and $16$ at TH wavelength, respectively, $\hat{\chi}^{(3)}$ is defined by the space group 227 (e.g. Si).}
    \label{fig:5}
\end{figure*}
%%%%%%%

We can apply this to analysis of specific harmonic processes. In the linear regime ($N=1$), the cross-polarized conversion is prohibited for $m>2$, thus incident LCP (RCP) light generates only LCP (RCP) harmonic signal in transmission. For SHG ($N=2$), only $m=3$ allows to fulfill the conservation law with $s_1=-1$ and any $s_2$. Only harmonic photons with the opposite polarization can be generated in transmission, as shown in Fig.~\ref{fig:5}A, e.g. incident LCP light generates only RCP transmitted harmonic signal and vise versa, in agreement with Eq.~\eqref{eq:const2}. For the third-harmonic generation (THG) ($N=3$), $m=2$ and $m=4$ are permitted. For $m=2$, the structure can generate photons of both polarizations in transmission for the given circularly-polarized incident beam with $s_1=-1$ and $s_2=-2$, while for $m=4$ only cross-polarized generation in transmission geometry is possible with $s_1=-1$ and any $s_2$, as shown in Fig.~\ref{fig:5}C, similar to the case with $N=2$, $m=3$. The selection rules for different $N$ and $m$ are summarized in Table~\ref{tb:2}.
\bgroup
\def\arraystretch{1.3}%  1 is the default, change whatever you need
\begin{table}[]
\centering
\caption{\label{tb:2} Selection rules for the reduced nonlinear $S$-matrix elements for chiral harmonic generation for structures with symmetry ${\rm C}_m$. The legend is \YES\YES all elements are allowed, \YES\NO only $t_{\rm RR},t_{\rm LL},r_{\rm LR},r_{\rm RL}$ are allowed, \NO\YES only $t_{\rm RL},t_{\rm LR},r_{\rm LL},r_{\rm RR}$ are allowed, \NO\NO all elements are zero.\\ }
\begin{tabular}{|p{0.05\linewidth}||>{\centering\arraybackslash}p{0.12\linewidth}|>{\centering\arraybackslash}p{0.12\linewidth}|>{\centering\arraybackslash}p{0.12\linewidth}|>{\centering\arraybackslash}p{0.12\linewidth}|>{\centering\arraybackslash}p{0.12\linewidth}|}
\hline  ${\rm C}_m$ & \multicolumn{4}{l}{harmonic order $N$}  &  \\  
\hline  
$m$
& 1    
& 2
& 3  
& 4
& 5  \\ \hline
1 
&    \YES\YES
&    \YES\YES
&    \YES\YES
&    \YES\YES
&    \YES\YES
\\
2 
&    \YES\YES
&    \NO\NO
&    \YES\YES
&    \NO\NO
&    \YES\YES
\\
3 
&    \YES\NO
&    \NO\YES
&    \NO\NO
&    \YES\NO
&    \NO\YES
\\
4 
&    \YES\NO
&    \NO\NO
&    \NO\YES
&    \NO\NO
&    \YES\NO
\\
6
&    \YES\NO
&    \NO\NO
&    \NO\NO
&    \NO\NO
&    \NO\YES
\\
\hline
\end{tabular}
\end{table}
\egroup
We confirm the developed theory with numerical calculations. More details about calculations can be found in Methods, see \ref{app:methods}. We first consider a dielectric metasurface made of the transition metal dichalcogenide MoS$_2$ in $3R$ phase, characterized with the C$_3$ symmetry of meta-atom, hexagonal metasurface lattice, and C$_3$ symmetry of the susceptibility tensor, shown in Fig.~\ref{fig:5}A. Therefore, for metasurface and susceptibility tensor lattices aligned, the symmetry of the nonlinear $S$-matrix is $m=3$. We consider the structure pumped with a circularly polarized plane wave propagating along the out-of-plane axis. Figure~\ref{fig:5}B shows the linear transmission, SH signal, linear CD and nonlinear CD in the near-infrared range of incident wavelengths. One can see that in the linear scattering, cross polarization conversion is prohibited, so RCP (LCP) light generates only RCP (LCP) transmitted signal. This agrees with the prediction made in the linear regime ($N=1$) for $m>2$ [Table~\ref{tb:2}]. The linear CD is nonzero and is governed by resonances of the structure. In the forward generated SH signal, the co-polarized components are zero as predicted by the selection rules [Table~\ref{tb:2}]. The nonlinear SH CD is large and is governed by resonances. 

Next, we consider a dielectric metasurface made of Si, characterized with the C$_4$ symmetry of meta-atom, square metasurface lattice, and cubic symmetry of the susceptibility tensor, shown in Fig.~\ref{fig:5}C. The symmetry of the nonlinear $S$-matrix is $m=4$. Figure~\ref{fig:5}D shows the linear transmission, SH signal, linear CD and nonlinear CD in the near-infrared range of incident wavelengths. Similar to Fig.~\ref{fig:5}B, in linear transmission the conversion of polarization is prohibited, while for nonlinear signal it is opposite and co-polarized third-harmonic (TH) signal is not generated in the forward direction. This confirms once again the prediction made by analysing the selection rules for the reduced nonlinear $S$-matrix.

These second-harmonic and third-harmonic examples give some insights that the reduced $S$-matrix with the simple selection rules can be helpful in designing a nonlinear chiral metasurface for experiments. Additionally, the knowledge of which circular polarization is expected to be transmitted or reflected at a given harmonic order will be significantly useful in constructing an optical measurement setup including different types of waveplates and polarizers.

\subsection{In-plane mirror symmetry}

\begin{figure}[t]
\centering
\includegraphics[width=0.9\linewidth]{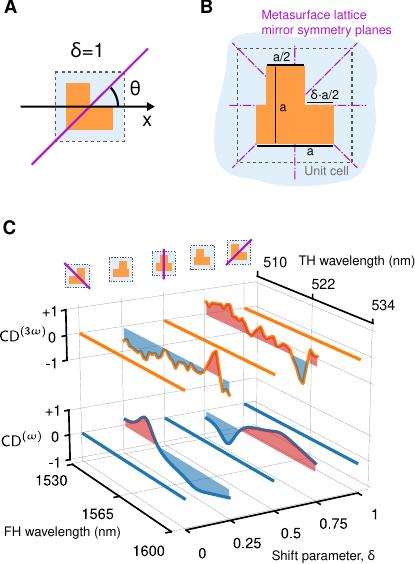}
\caption{{\bf Metasurfaces with in-plane mirror symmetries.} \boldsf{A} Definition of in-plane mirror symmetry plane (purple) and relative angle $\theta$. \boldsf{B} Schematic of metasurface inverse T-shaped meta-atom with shift parameter $\delta$ controlling the horizontal position of the upper bar. For $\delta=0$ the meta-atom has L-shape with two equal arms of length $a$. The meta-atoms are arranged in a square lattice. The metasurface lattice has 4 distinct in-plane mirror symmetry planes. For $\delta \neq 0, 0.5, 1$ the meta-atom does not have in-plane mirror symmetry planes, thus the metasurface as a whole does not have an in-plane mirror symmetry. \boldsf{C} Calculation of linear and TH CD. The pump wave is incident from the upper side. The CD vanishes for linear and nonlinear TH signal for $\delta = 0, 0.5, 1$ due to metasurface retaining one of in-plane mirror symmetries. Parameters of the simulation: meta-atom arm length $a = 900~\text{nm}$, height $h = 400~\text{nm}$, square lattice with period $1200~\text{nm}$, substrate permittivity $2$, material permittivity $16$, nonlinear tensor $\chi^{(3)}$ is defined by the space group 227 (e.g. Si).}
\label{fig:6}
\end{figure}

We next analyse metasurfaces with an in-plane mirror symmetry, defined with respect to an inclined in-plane axis rotated by the angle $\theta$ from the $x$-axis, as shown in Fig.~\ref{fig:6}A. We note that $\theta$ has no direct physical meaning, because $x,y$ axes can be redefined by rotating the coordinate system. The corresponding transformation matrix in the helical basis is
\begin{equation}
\hat{T}^{(\theta)}_{\pm}=\left(\begin{array}{cccc}
0 & 0 & -e^{2i\theta} & 0 \\
0 & 0  & 0 & -e^{2i\theta} \\
-e^{-2i\theta} & 0 & 0  & 0 \\
0 & -e^{-2i\theta} & 0 & 0 
\end{array}\right).
\end{equation}
The matrix is unitary, and $\left(\hat{T}^{(\theta)}_{\pm}\right)^{-1}=\hat{T}^{(\theta)}_{\pm}$. We can write the selection rules Eq.~\eqref{eq:SfinN} as,
\begin{equation}
    \hat{\bar{S}}^{(N\omega)} = (-1)^{N-1}\hat{T}^{(\theta)}_{\pm}\hat{\bar{S}}^{(N\omega)}\hat{T}^{(N\theta)}_{\pm}.
    \label{eq:Shelsig}
\end{equation}

%======================Figure============================
%======================Figure============================
Applying the selection rules to Eq.~\eqref{eq:Shelsig}, we get
% \begin{equation}
% \begin{aligned}
% &\frac{t_{\rm LL, u}^{(N)}}{t_{\rm RR, u}^{(N)}}=\frac{t_{\rm LL, d}^{(N)}}{t_{\rm RR, d}^{(N)}}=\left(-1\right)^{N-1}e^{-2i(N-1)\theta},\\
% &\frac{t_{\rm LR, u}^{(N)}}{t_{\rm RL, u}^{(N)}}=\frac{t_{\rm LR, d}^{(N)}}{t_{\rm LR, u}^{(N)}}=\left(-1\right)^{N+1}e^{-2i(N+1)\theta}.
% \end{aligned}
% \label{eq:alot}
% \end{equation}
% This means, that nonlinear OR is equal for excitation from both sides $p={\rm u,d}$ and is defined by the angle $\theta$ as 
% \begin{equation}
% \begin{aligned}
%     \mathrm{OR}^{(N)}_{ {\rm co}} &= (N-1)\left(\frac{\pi}{2}-\theta\right),\\
%     \mathrm{OR}^{(N)}_{{\rm cross}} &= (N+1)\left(\frac{\pi}{2}-\theta\right).
% \end{aligned}
% \end{equation}
% We can also conclude that for arbitrary $N$ and $\theta$,
\begin{equation}
\begin{aligned}
|t_{\rm LL, u(d)}^{(N)}|&=|t_{\rm RR, u(d)}^{(N)}|,\quad |t_{\rm LR, u(d)}^{(N)}|=|t_{\rm RL, u(d)}^{(N)}|,\\
|r_{\rm LL, u(d)}^{(N)}|&=|r_{\rm RR, u(d)}^{(N)}|,\quad |r_{\rm LR, u(d)}^{(N)}|=|r_{\rm RL, u(d)}^{(N)}|. 
\end{aligned}
\end{equation}
Therefore, the nonlinear CD in all definitions reaches zero for an arbitrary $N$ and $\theta$,
\begin{equation}
    \mathrm{CD}^{(N)}_{\rm co} = \mathrm{CD}^{(N)}_{\rm cross} = \mathrm{CD}^{(N)}_{\rm tot}=0.
\end{equation}
This is in agreement with the definition of geometrical chirality as absence of all mirror symmetries of the object. The selection rules for different $N$ are summarized in Table~\ref{tb:3}. 

%%%%%%%%
\begin{figure*}[t]
    \centering
    \includegraphics[width=0.93\linewidth]{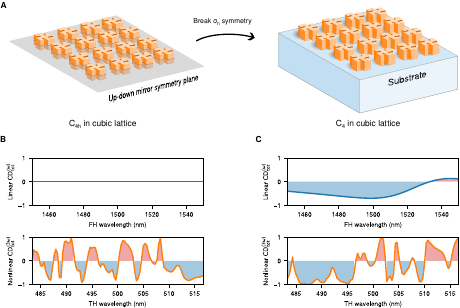}
    \caption{{\bf Metasurfaces with out-of plane mirror symmetry.} \boldsf{A} Illustration of the vertical mirror symmetry $\sigma_{\text{h}}$ plane and symmetry breaking scenario. (B,C)  Linear and TH nonlinear CD for a Si metasurface with $\mathrm{C}_{4\mathrm{h}}$ meta-atoms in a square lattice in the near-infrared range. The pump wave is incident from the upper side.  Parameters of the structure are the same as in the Fig.~\ref{fig:5}~C. \boldsf{B} Linear CD is prohibited by the combination of up-down symmetry and reciprocity. \boldsf{C} Both linear and nonlinear CD are allowed by symmetry.}
    \label{fig:7}
\end{figure*}
%%%%%%%%%%

% Similar relations can be written for the nonlinear reflection coefficients,
% \begin{equation}
% \begin{aligned}
% &\frac{r_{\rm LR, u}^{(N)}}{r_{\rm RL, u}^{(N)}}=\frac{r_{\rm LR, d}^{(N)}}{r_{\rm RL, d}^{(N)}}=\left(-1\right)^{N-1}e^{-2i(N-1)\theta},\\
% &\frac{r_{\rm LL, u}^{(N)}}{r_{\rm RR, u}^{(N)}}=\frac{r_{\rm LL, d}^{(N)}}{r_{\rm RR, d}^{(N)}}=\left(-1\right)^{N+1}e^{-2i(N+1)\theta},
% \end{aligned}
% \end{equation}
% which gives
% \begin{equation}
% \begin{aligned}
% &|r_{\rm LL, u(d)}^{(N)}|=|r_{\rm RR, u(d)}^{(N)}|,\\
% &|r_{\rm LR, u(d)}^{(N)}|=|r_{\rm RL, u(d)}^{(N)}|. 
% \end{aligned}
% \end{equation}

We confirm the developed theory with numerical calculations, for details see Methods in \ref{app:methods}. We consider a dielectric Si metasurface characterized with the shift parameter $\delta$ for the meta-atom, square metasurface lattice, and cubic symmetry of the susceptibility tensor, shown in Fig.~\ref{fig:6}B. Figure~\ref{fig:6}C shows linear CD and nonlinear TH CD for pump in the near-infrared range. One can see that for parameter $\delta=0,0.5,1$, both linear and nonlinear CD are zero, as expected, because the meta-atom and metasurface lattice mirror symmetry planes coincide and the structure becomes geometrically achiral.

\subsection{Out-of-plane mirror symmetry}

We finally analyse metasurfaces with the out-of-plane (up-down) mirror symmetry ($\sigma_{\rm h}$), defined with the matrix
\begin{equation}
\hat{T}^{(\sigma_{\rm h})}_{\pm}=\left(\begin{array}{cccc}
0 & 1 & 0 & 0 \\
1 & 0  & 0 & 0 \\
0 & 0 & 0 & 1 \\
0 & 0 & 1 & 0 
\end{array}\right).
\end{equation}
The matrix is unitary, and $\left(\hat{T}^{(\sigma_{\rm h})}_{\pm}\right)^{-1}=\hat{T}^{(\sigma_{\rm h})}_{\pm}$. We can write the selection rules Eq.~\ref{eq:SfinN} as
\begin{equation}
    \hat{\bar{S}}^{(N\omega)} = \hat{T}^{(\sigma_{\rm h})}_{\pm}\hat{\bar{S}}^{(N\omega)}\hat{T}^{(\sigma_{\rm h})}_{\pm}.
    \label{eq:Shelud}
\end{equation}
We conclude from Eq.~\ref{eq:Shelud} that the selection rules are independent on the harmonic order $N$. Therefore, we can write that the transmission and reflection amplitudes are equal on both sides of the metasurface upon changing from the left- to right- polarization and vise versa
\begin{equation}
\begin{aligned}
&t_{\rm RR, u(d)}^{(N)}=t_{\rm LL, d(u)}^{(N)},\quad t_{\rm LR, u(d)}^{(N)}=t_{\rm RL, d(u)}^{(N)},\\
&r_{\rm RR, u(d)}^{(N)}=r_{\rm LL, d(u)}^{(N)},\quad  r_{\rm LR, u(d)}^{(N)}=r_{\rm RL, d(u)}^{(N)}.
\end{aligned}
\end{equation}
Therefore, for all definitions of CD in Eq.~\eqref{eq:CDall},
\begin{equation}
\begin{aligned}
\mathrm{CD}^{(N)}_{\rm u}&=-\mathrm{CD}^{(N)}_{\rm d}.
\end{aligned}
\label{eq:onemore}
\end{equation}

\bgroup
\def\arraystretch{1.6}%  1 is the default, change whatever you need
\begin{table}[t]
\centering
\caption{\label{tb:3} Selection rules for CD for structures with in-plane mirror, out-of-plane mirror, and three-dimensional inversion symmetries. \\ }
\begin{tabular}{|p{0.38\linewidth}|>{\arraybackslash}p{0.23\linewidth}|>{\arraybackslash}p{0.23\linewidth}|}
\hline  in-plane & out-of-plane & 3D inversion \\  
\hline  
$\mathrm{CD}^{(1)}_{\rm co} = \mathrm{CD}^{(1)}_{\rm cross} = 0$ & $\mathrm{CD}^{(1)}_{ {\rm co}}=0$ &$\mathrm{CD}^{(1)}_{ {\rm co}}=0$  \\
$\mathrm{CD}^{(N)}_{\rm co}=0$ & $\mathrm{CD}_{ {\rm co}}^{(N)}\ne0$ & $\mathrm{CD}_{ {\rm co}}^{(N)}\ne0$ \\
$\mathrm{CD}^{(N)}_{\rm cross}=0$ & $\mathrm{CD}_{\rm cross}^{(N)}\ne0$ & $\mathrm{CD}_{\rm cross}^{(N)}\ne0$ \\\hline
\end{tabular}
\end{table}
\egroup

\paragraph{Linear scattering and reciprocity.} For linear processes ($N=1$), the fundamental principle of the Lorentz reciprocity adds additional constraints on the $S$-matrix. The selection rules imposed by the reciprocity can be written in the helical basis as
\begin{equation}
    \hat{\bar{S}}^{(\omega)} = \hat{T}^{({\rm R})}_{\pm}\left(\hat{\bar{S}}^{(\omega)}\right)^{\rm T}\hat{T}^{({\rm R})}_{\pm},
    \label{eq:rechel}
\end{equation}
where $\hat{T}^{({\rm R})}_{\pm}$ is the corresponding transformation matrix (see \ref{app:3}). The $S$-matrix coefficients are 
\begin{equation}
\begin{aligned}
&t_{\rm RR, u}^{(1)}=t_{\rm RR, d}^{(1)},\quad t_{\rm LL, u}^{(1)}=t_{\rm LL, d}^{(1)},\\
&t_{\rm RL, u}^{(1)}=t_{\rm LR, d}^{(1)},\quad t_{\rm RL, d}^{(1)}=t_{\rm LR, u}^{(1)},\\
&r_{\rm RL, u}^{(1)}=r_{\rm LR, u}^{(1)},\quad r_{\rm RL, d}^{(1)}=r_{\rm LR, d}^{(1)}.
\label{eq:51}
\end{aligned}
\end{equation}
Applying Eq.~\eqref{eq:51}, we find that the co-polarized linear CD is zero, independently of the excitation side $p={\rm u,d}$,
\begin{equation}
\begin{aligned}
&\mathrm{CD}^{(1)}_{\rm co}=0,\\
&\mathrm{CD}^{(1)}_{\rm tot}=\mathrm{CD}^{(1)}_{\rm cross}.
\end{aligned}
\label{eq:andonemore}
\end{equation}
Thus, we can design structures with out-of-plane mirror symmetry, that demonstrate a zero linear CD reaching, but nonzero nonlinear CD, in case the in-plane mirror symmetries are not present.  The selection rules for different $N$ are summarized in Table~\ref{tb:3}. 

We confirm the developed theory with numerical calculations, for details see Methods in \ref{app:methods}. We consider the same dielectric Si metasurface as in Fig.~\ref{fig:5}C, with ${\rm C}_4$ rotational symmetry. We compare the linear and nonlinear TH signal for the metasurface with and without the substrate, as shown in Fig.~\ref{fig:7}A. Figures~\ref{fig:7}B,C show that the linear CD in transmission vanishes in the absence of the substrate, as expected from the reciprocity and symmetry considerations, while the nonlinear CD is non-zero. 

\subsection{Other symmetries}
\paragraph{Inversion symmetry.} We assume that the structure has the three-dimensional inversion symmetry ${\rm C_i}$ that represents a product of ${\rm C}_2$ and $\sigma_{\rm h}$. Then, the transformation matrix is
\begin{equation}
\hat{T}^{({\rm I})}_{\pm}=\hat{T}^{(\varphi_2)}_{\pm}\hat{T}^{(\sigma_{\rm h})}_{\pm}=\left(\begin{array}{cccc}
0 & -1 & 0 & 0 \\
-1 & 0  & 0 & 0 \\
0 & 0 & 0 & -1 \\
0 & 0 & -1 & 0 
\end{array}\right).
\end{equation}
The matrix is unitary, and $\left(\hat{T}^{({\rm I})}_{\pm}\right)^{-1}=\hat{T}^{({\rm I})}_{\pm}$. We can write the selection rules Eq.~\eqref{eq:SfinN} as
\begin{equation}
    \hat{\bar{S}}^{(N\omega)} = (-1)^{(N-1)}\hat{T}^{({\rm I})}_{\pm}\hat{\bar{S}}^{(N\omega)}\hat{T}^{({\rm I})}_{\pm}.
    \label{eq:ShelSm}
\end{equation}
% \bgroup
% \def\arraystretch{1.4}%  1 is the default, change whatever you need
% \begin{table}[]
% \centering
% \caption{\label{tb:4} Selection rules for the reduced nonlinear $S$-matrix elements for chiral harmonic generation for structures with symmetry ${\rm S}_m$ vs. $m$ and $N$. The legend is \YES\YES all elements are allowed, \YES\NO only $t_{\rm RR},t_{\rm LL},r_{\rm LR},r_{\rm RL}$ are allowed, \NO\YES only $t_{\rm RL},t_{\rm LR},r_{\rm LL},r_{\rm RR}$ are allowed, \NO\NO all elements are zero.\\}
% \begin{tabular}{|p{0.05\linewidth}||>{\centering\arraybackslash}p{0.12\linewidth}|>{\centering\arraybackslash}p{0.12\linewidth}|>{\centering\arraybackslash}p{0.12\linewidth}|>{\centering\arraybackslash}p{0.12\linewidth}|>{\centering\arraybackslash}p{0.12\linewidth}|}
% \hline  ${\rm S}_m$ & \multicolumn{4}{l}{harmonic order $N$}  &  \\  
% \hline
% $m$
% & 1    
% & 2
% & 3  
% & 4
% & 5  \\ \hline
% %
% 1 
% &    \YES\YES
% &    \YES\YES
% &    \YES\YES
% &    \YES\YES
% &    \YES\YES
% \\
% %
% %
% 2 
% &    \YES\YES
% &    \YES\YES
% &    \YES\YES
% &    \YES\YES
% &    \YES\YES
% \\
% %
% %
% 3 
% &    \YES\NO
% &    \NO\YES
% &    \NO\NO
% &    \YES\NO
% &    \NO\YES
% \\
% %
% %
% 4 
% &    \YES\YES
% &    \NO\NO
% &    \YES\YES
% &    \NO\NO
% &    \YES\YES
% \\
% \hline
% %
% \end{tabular}
% \end{table}
% \egroup
Applying the selection rules to Eq.~(\ref{eq:ShelSm}), we get
\begin{equation}
\begin{aligned}
&\frac{t_{\rm RR, u(d)}^{(N)}}{t_{\rm LL, d(u)}^{(N)}}=\frac{t_{\rm LR, u(d)}^{(N)}}{t_{\rm RL, d(u)}^{(N)}}=(-1)^{(N-1)},\\
&\frac{r_{\rm RR, u(d)}^{(N)}}{r_{\rm LL, d(u)}^{(N)}}=\frac{r_{\rm LR, u(d)}^{(N)}}{r_{\rm RL, d(u)}^{(N)}}=(-1)^{(N-1)}.
\end{aligned}
\end{equation}
Thus, generation of both co- and cross-polarized chiral harmonics is allowed for circularly polarized incident beam, and Eq.~\eqref{eq:onemore} is fulfilled. For $N=1$, Eqs.~\eqref{eq:51} and~\eqref{eq:andonemore} are valid due to the reciprocity.  The selection rules for different $N$ are summarized in Table~\ref{tb:3}. 

\section{Conclusion and outlook}

We have demonstrated that a large class of nonlinear frequency mixing processes in planar metasurfaces can be described within the framework of the scattering matrix approach similar to the case of the conventional linear scattering processes. We have introduced the definition and described properties of the nonlinear scattering matrix. We have shown that for optical harmonic generation the scattering matrix can be reduced to a simpler two-dimensional form. We have derived the simplified selection rules for the reduced scattering matrix in the case of metasurfaces obeying specific point group symmetry transformations, and have applied them to the analysis of chiral harmonic generation in nonlinear metasurfaces with rotational, mirror reflection and inversion symmetries, see Tables~\ref{tb:2} and~\ref{tb:3}. We have verified our analytical results with numerical calculations.
 
We note that the developed framework is frequency independent, as it relies on symmetries only, and can be applied to analysis of both resonant and non-resonant nonlinear response. Moreover, the method can be applied for nonlinear metasurfaces with misaligned microscopic and macroscopic lattices, as long as each of point symmetry groups (meta-atom geometry, metasurface lattice arrangement and susceptibility tensor) includes the specific symmetry element. 

We emphasize that the presented approach can be generalized to describe a variety of nonlinear frequency mixing processes, including but not limited to difference frequency generation, four-wave mixing, and self-induced nonlinear effects. The described formalism can be also generalized to the analysis of oblique incidence and analysis of diffraction effects with a larger number of scattering channels.

The presented nonlinear scattering matrix approach is not only theoretically insightful and robust but also has practical usage. Nonlinear chiral metasurfaces can be designed considering the selection rules at the desired harmonics prior to performing full-wave simulations and fabrication of samples. The material and meta-atom symmetries as well as the lattice structure and the substrate can be selected for a desired function of the metasurface. This provides a well-established engineering guideline strongly supported by solid theory. We believe our results can find many important applications for engineering nonlinear metasurfaces with the properties on demand for polarization control, filtering, and nonlinear wave-front shaping.

% \clearpage

\appendix
\section{Nonlinear $S$-matrix derivation from Maxwell's equations}
\label{app:1}
In this section we derive an expression for the nonlinear $S$-matrix $\hat{S}^{(\omega_1+\ldots+\omega_N)}$ explicitly from the Maxwell equations. We will show that the nonlinear $S$-matrix obeys Eq.~(\ref{eq:Str}) only if the specific point group symmetry $\hat{T}$ is achieved both at the macroscopic scale for the metasurface lattice and meta-atom, and microscopic scale for the linear permittivity and nonlinear susceptibility tensor.

We consider a planar metasurface pumped with $N$ multiple frequencies $\omega_1,\ldots,\omega_N$ that generates a signal at the output frequency $\omega_{\rm out}=\omega_1+\ldots+\omega_N$. We assume the perturbative regime and undepleted pump approximation for the frequency conversion. The induced nonlinear polarization at $\omega_{\rm out}$ is
\begin{equation}
P_{i_0}^{(\omega_{\rm out})}=\sum\limits_{i_1,\ldots, i_N} \chi_{i_0i_1\ldots i_N}^{(N)} E_{i_1}(\omega_1) \cdots E_{i_N}(\omega_N).
\label{eq:Pnlexp}
\end{equation}
Here, $i_0,i_1\ldots,i_N=x,y,z$, and $\mathbf{E}(\omega_l)$ is the full field at the corresponding frequency $\omega_l$ of the input beam $l=1\ldots N$, given by
\begin{multline}
\mathbf{E}(\mathbf{r} ; \omega_l)=\mathbf{E}_{\mathrm{bg}}\left(\mathbf{r};  \omega_l\right)-k_l^2 \int \dd \mathbf{r}^{\prime}\  \hat{\mathbf{G}}_{\mathrm{EE}}\left(\mathbf{r}, \mathbf{r}^{\prime} ; \omega_l\right) \cdot \\ \cdot\left[\varepsilon\left(\mathbf{r}^{\prime}\right)-\varepsilon_{\rm bg}\left(\mathbf{r}^{\prime}\right)\right]\mathbf{E}_{\mathrm{bg}}\left(\mathbf{r}^{\prime} ;  \omega_l\right),
\label{eq:E1exp}
\end{multline}
where $k_l=\omega_l/c$, $\varepsilon$ is the metasurface permittivity, $\varepsilon_{\rm bg}$ is the background permittivity, $\mathbf{E}_{\mathrm{bg}}(\omega_l)$ is the incident background field at the input frequency $\omega_l$ that satisfies
\begin{equation}
\nabla \times \nabla \times \mathbf{E}_{\mathrm{bg}}(\mathbf{r})=k_l^2 \varepsilon_{\rm bg}(\mathbf{r}) \mathbf{E}_{\mathrm{bg}}(\mathbf{r}).
\end{equation}
Here, $\hat{\mathbf{G}}_{\mathrm{EE}}$ is the electric-electric Green function of the resonant metasurface at frequency $\omega_l$, defined as
\begin{multline}
-\nabla \times \nabla \times \hat{\mathbf{G}}_{\mathrm{EE}}\left(\mathbf{r}, \mathbf{r}^{\prime} ; \omega_l\right)+\\
+k_l^2 \varepsilon(\mathbf{r}) \hat{\mathbf{G}}_{\mathrm{EE}}\left(\mathbf{r}, \mathbf{r}^{\prime} ; \omega_l\right)=\hat{\mathbf{1}} \delta\left(\mathbf{r}-\mathbf{r}^{\prime}\right).
\end{multline}

We are interested in the frequency of the input and nonlinear output beams below the diffraction order, thus we consider $4$ input and $4$ output channels. We define the $S$-matrix scattering channels as plane waves with specific polarization that carry power towards and away from the metasurface. As an example, we consider linearly polarized plane waves with polarizations along $x$ and $y$ axis, respectively. The input channels are at the input frequencies, and the output channels are at the output sum frequency. We define the electric fields in the channels as
\begin{equation}
\begin{aligned}
\mathbf{E}_{1}^{\mathrm{in}}&\equiv\mathbf{E}_{x,{\rm t}}^{\mathrm{in}}(\omega_l) =\sqrt{2} \hat{\mathbf{E}}_{x} e^{ i z k_l} H_\theta(-z), \\
\mathbf{E}_{2}^{\mathrm{in}}&\equiv\mathbf{E}_{y,{\rm t}}^{\mathrm{in}}(\omega_l) =\sqrt{2} \hat{\mathbf{E}}_{y} e^{ i z k_l} H_\theta(-z), \\
\mathbf{E}_{3}^{\mathrm{in}}&\equiv\mathbf{E}_{x,{\rm b}}^{\mathrm{in}}(\omega_l) =\sqrt{2} \hat{\mathbf{E}}_{x} e^{ -i z k_l} H_\theta(z),\\
\mathbf{E}_{4}^{\mathrm{in}}&\equiv\mathbf{E}_{y,{\rm b}}^{\mathrm{in}}(\omega_l) =\sqrt{2} \hat{\mathbf{E}}_{y} e^{ -i z k_l} H_\theta(z),\\
\mathbf{E}_{1}^{\mathrm{out}}&\equiv\mathbf{E}_{x,{\rm t}}^{\mathrm{out}}(\omega_{\mathrm{out}}) =\sqrt{2} \hat{\mathbf{E}}_{x} e^{ -i z k_{\mathrm{out}}} H_\theta(-z), \\
\mathbf{E}_{2}^{\mathrm{out}}&\equiv\mathbf{E}_{y,{\rm t}}^{\mathrm{out}}(\omega_{\mathrm{out}}) =\sqrt{2} \hat{\mathbf{E}}_{y} e^{ -i z k_{\mathrm{out}}} H_\theta(-z), \\
\mathbf{E}_{3}^{\mathrm{out}}&\equiv\mathbf{E}_{x,{\rm b}}^{\mathrm{out}}(\omega_{\mathrm{out}}) =\sqrt{2} \hat{\mathbf{E}}_{x} e^{ i z k_{\mathrm{out}}} H_\theta(z), \\
\mathbf{E}_{4}^{\mathrm{out}}&\equiv\mathbf{E}_{y,{\rm b}}^{\mathrm{out}}(\omega_{\mathrm{out}}) =\sqrt{2} \hat{\mathbf{E}}_{y} e^{ i z k_{\mathrm{out}}} H_\theta(z),
\end{aligned}
\end{equation}
where $\hat{\mathbf{E}}_{x(y)}$ are the unitary polarization vectors, the waves are propagating from the up ($z<0$) and bottom ($z>0$) direction, axis z is defined as in Fig.~\ref{fig:4}B, and $H_\theta(\pm z)$ is the Heaviside theta function. The magnetic fields can be reconstructed from the Maxwell equations. We assumed the fields in each channels are normalized such as they have a unity outgoing/incoming power, with their inner product defined via the Poynting flux.

We then define auxiliary functions $\mathbf{E}_{1\ldots4}^{\mathrm{inc}}(\omega_l)$ as propagating waves composed of incoming and outgoing fields that are free of singularities at the input frequency $\omega_l$,~\cite{zhang2020quasinormal}
\begin{equation}
\begin{aligned}
\mathbf{E}_{1}^{\mathrm{inc}} & =\mathbf{E}_{x,{\rm t}}^{\mathrm{in}}(\omega_l) + \mathbf{E}_{x,{\rm b}}^{\mathrm{out}}(\omega_l) =  \sqrt{2} \hat{\mathbf{E}}_{x} e^{ i z k_l},\\
\mathbf{E}_{2}^{\mathrm{inc}} & =\mathbf{E}_{y,{\rm t}}^{\mathrm{in}}(\omega_l) + \mathbf{E}_{y,{\rm b}}^{\mathrm{out}}(\omega_l) =  \sqrt{2} \hat{\mathbf{E}}_{y} e^{ i z k_l},\\
\mathbf{E}_{3}^{\mathrm{inc}} & =\mathbf{E}_{x,{\rm b}}^{\mathrm{in}}(\omega_l) + \mathbf{E}_{x,{\rm t}}^{\mathrm{out}}(\omega_l) =  \sqrt{2} \hat{\mathbf{E}}_{x} e^{ -i z k_l},\\
\mathbf{E}_{4}^{\mathrm{inc}} & =\mathbf{E}_{y,{\rm b}}^{\mathrm{in}}(\omega_l) + \mathbf{E}_{y,{\rm t}}^{\mathrm{out}}(\omega_l) =  \sqrt{2} \hat{\mathbf{E}}_{y} e^{ -i z k_l}.
\end{aligned}
\label{eq:Einc}
\end{equation}

Any incident field can be expanded over input and output amplitudes as
\begin{equation}
    \mathbf{E}_{\mathrm{bg}}=\sum\limits_{i=1}^4\left(a^{\mathrm{in}}_i(\omega_l)\mathbf{E}_{i}^{\mathrm{in}}+a^{\mathrm{out}}_i(\omega_l)\mathbf{E}_{i}^{\mathrm{out}}\right).
\end{equation}
At the same time, from Eq.~\eqref{eq:Einc}
\begin{equation}
    \mathbf{E}_{\mathrm{bg}}=\sum\limits_{i=1}^4a^{\mathrm{inc}}_i(\omega_l)\mathbf{E}_{i}^{\mathrm{inc}}.
    \label{eq:Ebgexp}
\end{equation}
Thus, $a^{\mathrm{inc}}_i=a^{\mathrm{in}}_i$, and we can define the incoming amplitude for nonlinear $S$-matrix as
\begin{equation}
    a_i(\omega_l)\equiv a^{\mathrm{in}}_i(\omega_l), \quad i=1\ldots4.
    \label{eq:aeqa}
\end{equation}

The nonlinear signal at the output frequency $\omega_{\rm out}$ can be found as,
\begin{equation}
\begin{aligned}
& \mathbf{E}(\omega_{\rm out})=-4\pi k_{\rm out}^2 \int \dd \mathbf{r}^{\prime} \hat{\mathbf{G}}_{\mathrm{EE}}( \omega_{\rm out}) \cdot \mathbf{P}^{(\omega_{\rm out})}, \\
& \mathbf{H}(\omega_{\rm out})=-4\pi k_{\rm out}^2 \int \dd \mathbf{r}^{\prime} \hat{\mathbf{G}}_{\mathrm{HE}}( \omega_{\rm out}) \cdot \mathbf{P}^{(\omega_{\rm out})},
\end{aligned}
\label{eq:NLexp}
\end{equation}
where the magnetic-electric Green function is $\hat{\mathbf{G}}_{\mathrm{HE}}=\nabla \times \hat{\mathbf{G}}_{\mathrm{EE}}/k_{\rm out}$.

The output signal can be expanded over channels with amplitudes $b_i(\omega_{\rm out})$ as
\begin{equation}
\begin{aligned}
& \mathbf{E}(\omega_{\rm out})=\sum\limits_{i=1}^4b_i(\omega_{\rm out})\mathbf{E}_{i}^{\mathrm{out}}(\omega_{\rm out}), \\
& \mathbf{H}(\omega_{\rm out})=\sum\limits_{i=1}^4b_i(\omega_{\rm out})\mathbf{H}_{i}^{\mathrm{out}}(\omega_{\rm out}),
\end{aligned}
\end{equation}
where
\begin{multline}
    b_i(\omega_{\rm out}) = \oint \dd \mathbf{S}\cdot\left[\left(\mathbf{E}_{i}^{\mathrm{out}}\right)^*\times\mathbf{H}(\omega_{\rm out})-\right.\\
    -\left.\left(\mathbf{H}_{i}^{\mathrm{out}}\right)^*\times\mathbf{E}(\omega_{\rm out})  \right].
    \label{eq:bexp}
\end{multline}

Combining Eqs.~\eqref{eq:Pnlexp},~\eqref{eq:E1exp},~\eqref{eq:Ebgexp},~\eqref{eq:aeqa},~\eqref{eq:NLexp}, and \eqref{eq:bexp}, we can introduce the nonlinear $S$-matrix as
\begin{equation}
b_i^{(\omega_{\rm out})}=\sum\limits_{j,\ldots,k}S^{(\omega_{\rm out})}_{ij\dots k}a_j^{(\omega_1)}\cdot\ldots\cdot a_k^{(\omega_N)}.
\end{equation}
Here, the $S$-matrix is defined as
\begin{multline}
S^{(\omega_{\rm out})}_{ij\dots k} =  -4\pi k_{\rm out}^2\sum\limits_{i_0,\ldots, i_N} \int \dd \mathbf{r}\ \chi_{i_0i_1\ldots i_N}^{(N)} \cdot\\
\cdot e_{i_{0}}^{(i)}(\omega_{\rm out})e_{i_{1}}^{(j)}(\omega_{1})\cdots e_{i_N}^{(k)}({\omega_{N}}),
\label{eq:appSfin}
\end{multline}
where $i_0,i_1,\ldots i_N=x,y,z$; $i,j\ldots k=1,..,4$. The integral in Eq.~\eqref{eq:appSfin} is the volume overlap of the susceptibility tensor, the output field,
\begin{multline}
\mathbf{e}^{(i)}(\omega_{\rm out})=  \oint \dd \mathbf{S}^\prime\cdot\left[\left(\mathbf{E}_{i}^{\mathrm{out}}\right)^*\times\hat{\mathbf{G}}_{\mathrm{HE}}(\omega_{\rm out})-\right.\\
-\left.\left(\mathbf{H}_{i}^{\mathrm{out}}\right)^*\times\hat{\mathbf{G}}_{\mathrm{EE}}(\omega_{\rm out})  \right],
\label{eq:fineout}
\end{multline}
and $N$ the input fields, with the field at $\omega_l$ ($l=1,\ldots,N$) given by
\begin{multline}
\mathbf{e}^{(i)}(\omega_l)= \mathbf{E}_{i}^{\mathrm{inc}}(\omega_l)-\\
-k_l^2 \int \dd \mathbf{r}^\prime\ \left[\varepsilon-\varepsilon_{\rm bg}\right]\hat{\mathbf{G}}_{\mathrm{EE}}\left(\omega_l\right) \cdot  
\mathbf{E}^{\mathrm{inc}}_i(\omega_l).
\label{eq:finein}
\end{multline}

We can identify the nonlinear $S$-matrix symmetry using Eqs.~\eqref{eq:appSfin},~\eqref{eq:finein} and \eqref{eq:fineout}. The symmetry of nonlinear $S$-matrix is defined by the microscopic symmetry via the permittivity $\varepsilon$ and the nonlinear susceptibility tensor $\hat{\chi}^{(N)}$, and by the macroscopic symmetry via the metasurface lattice and meta-atom symmetry (identified by the Green functions).

\section{Auxiliary $S$-matrix for arbitrary harmonic order}
\label{app:2}
In this section we introduce an auxiliary $S$-matrix  $\hat{\mathbb{S}}^{(N\omega)}$ for description of harmonic generation for an arbitrary harmonic number $N$, similar to Eq.~(\ref{eq:Saux2}). The definition of the auxiliary $S$-matrix is
\begin{equation}
    {\mathbb{S}}^{(N\omega)}_{i\alpha}\equiv\frac{{S}^{(N\omega)}_{ijk\ldots l}+{S}^{(N\omega)}_{ikj\ldots l}+\ldots+{S}^{(N\omega)}_{ilk\ldots j}}{N!},
\end{equation}
where $\hat{\mathbb{S}}^{(N\omega)}$ is a matrix of the second rank with the shape of ($d\times\alpha_{\rm max}$), $\alpha(j,k,\ldots,l)=1,\ldots,\alpha_{\rm max}$, and $d$ is the number of radiation channels (we used $d=4$ in the main text). The number of columns $\alpha_{\rm max}$ can be calculated as
\begin{equation}
    \alpha_{\rm max} = \frac{d(d+1)\ldots(d+N-1)}{N!}.
\end{equation}
The auxiliary symmetry transformation operator $\hat{\mathbb{T}}^{(N)}$  matrix of ($\alpha_{\rm max}\times\alpha_{\rm max}$) shape can be defined as
\begin{equation}
    \mathbb{T}^{(N)}_{\alpha\beta}\equiv\frac{T_{jp}T_{ks}+\ldots+T_{js}T_{kp}}{N!}.
\end{equation}

\section{Transformation matrices in Cartesian basis}
\label{app:3}
In this section we provide transformation matrices $\hat{T}_{xy}$ in the in the Cartesian basis of amplitudes $a_{x,y,{\rm u(d)}}$ and $b_{x,y,{\rm u(d)}}$ for waves travelling in positive and negative direction along the $z$ axis. The matrices can be constructed from analysis Eq.~\eqref{eq:ab}. The matrix corresponding to the rotational symmetry of $m$-th order (${\rm C}_m$) is
\begin{equation}
\hat{T}^{(\varphi_m)}_{xy}=\left(\begin{array}{cccc}
\cos{\varphi_m} & 0 & \sin{\varphi_m} & 0 \\
0 & \cos{\varphi_m}  & 0 & \sin{\varphi_m} \\
-\sin{\varphi_m} & 0 & \cos{\varphi_m}  & 0 \\
0 & -\sin{\varphi_m} & 0 & \cos{\varphi_m} 
\end{array}\right).
\end{equation}
The matrix for the in-plane mirror symmetry, defined with respect to an inclined axis rotated by the angle $\theta$ from the $x$-axis, is
\begin{equation}
\hat{T}^{(\theta)}_{xy}=\left(\begin{array}{cccc}
-\cos{(2\theta)} & 0 & \sin{\left(2\theta\right)} & 0 \\
0 & -\cos{(2\theta)}  & 0 & \sin{\left(2\theta\right)} \\
\sin{\left(2\theta\right)} & 0 & \cos{(2\theta)}  & 0 \\
0 & \sin{\left(2\theta\right)} & 0 & \cos{(2\theta)} 
\end{array}\right).
\end{equation}
The matrix corresponding to out-of-plane up-down mirror symmetry ($\sigma_{\rm h}$) is
\begin{equation}
\hat{T}^{(\sigma_{\rm h})}_{xy}=\left(\begin{array}{cccc}
0 & 1 & 0 & 0 \\
1 & 0  & 0 & 0 \\
0 & 0 & 0 & 1 \\
0 & 0 & 1 & 0 
\end{array}\right).
\end{equation}
One can see that matrices $\hat{T}^{(\varphi_m)}_{xy}$ and $\hat{T}^{(\theta)}_{xy}$ do not obey Eq.~\eqref{eq:Tcond}, thus the reduced nonlinear $S$-matrix approach cannot be applied for these transformations.

To calculate the corresponding matrices in the helical basis $\hat{T}_{\pm}$, we use matrices $\hat{T}_{xy}$ and unitary transformation
\begin{equation}
\hat{T}_{\pm}=\hat{T}_{\rm c} \hat{T}_{x y} \hat{T}_{\rm c}^{\dagger},
\end{equation}
where $\hat{T}_{\rm c}$ is the basis transformation matrix
\begin{equation}
\hat{T}_{\rm c}=\frac{1}{\sqrt{2}}\left(\begin{array}{cccc}
1 & 0 & -i & 0 \\
0 & 1 & 0 & -i \\
1 & 0 & i & 0 \\
0 & 1 & 0 & i
\end{array}\right).
\end{equation}

The selection rules for the linear $S$-matrix $\hat{\bar{S}}^{(\omega)}_{xy}$ imposed by the Lorentz reciprocity can be written in Cartezian basis as 
\begin{equation}
    \hat{\bar{S}}^{(\omega)}_{xy} = \left(\hat{\bar{S}}^{(\omega)}_{xy}\right)^{\rm T}.
\end{equation}
To transform them into the helical basis, we apply the unitary transformation to the $S$-matrix,
\begin{equation}
\hat{\bar{S}}^{(\omega)}=\hat{T}_{\rm c} \hat{\bar{S}}_{xy}^{(\omega)} \hat{T}_{\rm c}^{\dagger}.
\end{equation}
This gives us Eq.~\eqref{eq:rechel} with $\hat{T}^{({\rm R})}_{\pm}$ defined as
\begin{equation}
\hat{T}^{({\rm R})}_{\pm}=\hat{T}_{\rm c}\hat{T}_{\rm c}^{\rm T}=\left(\begin{array}{cccc}
0 & 0 & 1 & 0 \\
0 & 0  & 0 & 1 \\
1 & 0 & 0 & 0 \\
0 & 1 & 0 & 0 
\end{array}\right).
\end{equation}

\section{Methods: numerical calculations.}
\label{app:methods}
For numerical simulations of the transmittance and harmonic spectra, we use the finite-element-method solver in COMSOL Multiphysics in the frequency domain. All calculations were realized for a metasurface placed on a semi-infinite substrate surrounded by a perfectly matched layer mimicking an infinite region. The simulation area is the unit cell extended to an infinite metasurface by using the Bloch boundary conditions. The material properties are listed in the corresponding figure caption. The incident field is a plane wave in the normal excitation geometry with left or right circular polarization incident from the upper side of the metasurface. 

The nonlinear simulations of chiral harmonic response are employed within the undepleted pump approximation, using two steps to calculate the intensity of the radiated nonlinear signal. First, we simulated the linear scattering at the pump wavelength, and then obtained the nonlinear polarization induced inside the meta-atom. Then, we employed the polarization as a source for the electromagnetic simulation at the harmonic wavelength to obtain the generated harmonic field. The nonlinear susceptibility $\hat{\chi}^{(N)}$ was considered as a constant tensor corresponding to the specific crystallographic point group listed in the corresponding figure caption. 

\ack
K. Koshelev acknowledges Maxim Gorkunov and Kristina Frizyuk for many useful discussions and suggestions. This work was supported by the Australian Research Council (Grant
DP210101292) and the International Technology Center Indo-Pacific (ITC IPAC) via Army Research Office (contract FA520923C0023). The authors have no relevant financial interests in the paper and no other potential conflicts of interest to disclose.

K.K. conceived the original idea and supervised the project. I.T. performed numerical calculations and modelling. All authors discussed, interpreted, and corroborated the results. All authors participated in the preparation and writing of the manuscript. 

\section*{References}

\bibliography{refs}

\end{document}